\documentclass[%
 reprint,
superscriptaddress,
 amsmath,amssymb,
 aps,
]{revtex4-2}

\usepackage{graphicx}
\graphicspath{ {./images/} }
\usepackage{siunitx}
\usepackage[dvipsnames]{xcolor}
\usepackage{bm}%
\usepackage{hyperref}%

\begin{document}

\newcommand{\kBT}{k_\mathrm{B}T}
\newcommand{\tm}{\mathrm}
\newcommand{\Vb}{V}
\newcommand{\nuc}{\nu_\mathrm{c}}
\newcommand{\nub}{\nu_\mathrm{b}}
\newcommand{\nus}{\nu_\mathrm{s}}
\newcommand{\muc}{\mu_\mathrm{c}}
\newcommand{\mub}{\mu_\mathrm{b}}
\newcommand{\phis}{\phi}
\newcommand{\phib}{\psi}
\newcommand{\Jint}{J_\mathrm{int}}
\newcommand{\Jdiff}{J_\mathrm{diff}}

\newcommand{\phic}{\phi_\mathrm{c}}

\newcommand{\phizin}{\phi_\mathrm{in}^{(0)}}
\newcommand{\phizout}{\phi_\mathrm{out}^{(0)}}
\newcommand{\phizn}{\phi_{n}^{(0)}}

\newcommand{\cs}{c_\mathrm{s}}
\newcommand{\cb}{c_\mathrm{b}}
\newcommand{\mus}{\mu_\mathrm{s}}
\newcommand{\cin}{c_\mathrm{in}^\mathrm{eq}}
\newcommand{\cout}{c_\mathrm{out}^\mathrm{eq}}

\newcommand{\ka}{k_\mathrm{a}}
\newcommand{\kp}{k_\mathrm{p}}
\newcommand{\ellc}{l_\mathrm{c}}

\newcommand{\vect}{\boldsymbol}
\newcommand{\diff}{\mathrm{d}}
\newcommand{\PhiTot}{\Phi_{\mathrm{tot}}}
\newcommand{\Domega}{\Delta\omega}
\newcommand{\DomegaEff}{\Domega_\mathrm{eff}}
\newcommand{\Asys}{A_\mathrm{sys}}

\newcommand{\figref}[1]{Fig.~\ref{#1}}
\newcommand{\appref}[1]{SI.~\ref{#1}}
\newcommand{\secref}[1]{Sec.~\ref{#1}}
\newcommand{\Eqref}[1]{Eq.~\eqref{#1}}
\newcommand{\Eqsref}[1]{Eqs.~\eqref{#1}}
\newcommand{\Kd}{K_\mathrm{d}}
\newcommand{\cd}{c_\mathrm{d}}
\newcommand{\DZ}[1]{\textcolor{purple}{DZ: #1}}
\newcommand{\RR}[1]{\textcolor{teal}{RR: #1}}
\newcommand{\ME}[1]{\textcolor{olive}{ME: #1}}
\newcommand{\olig}{\ell}

\let\oldsection\section
\let\oldsubsection\subsection

\renewcommand{\tableofcontents}{}
\renewcommand{\section}[1]{}
\renewcommand{\subsection}[1]{}

\preprint{APS/123-QED}

\title{Exchange controls coarsening of surface condensates}

\author{Riccardo Rossetto}
\affiliation{Max Planck Institute for Dynamics and Self-Organization, Am Faßberg 17, 37077 Göttingen, Germany}%
\affiliation{University of Göttingen, Institute for the Dynamics of Complex Systems, Friedrich-Hund-Platz 1, 37077 Göttingen, Germany}%
\author{Marcel Ernst}
\affiliation{Max Planck Institute for Dynamics and Self-Organization, Am Faßberg 17, 37077 Göttingen, Germany}%
\affiliation{University of Göttingen, Institute for the Dynamics of Complex Systems, Friedrich-Hund-Platz 1, 37077 Göttingen, Germany}%
\author{David Zwicker}%
 \email{david.zwicker@ds.mpg.de}
\affiliation{Max Planck Institute for Dynamics and Self-Organization, Am Faßberg 17, 37077 Göttingen, Germany}%

\begin{abstract}

Biological membranes often exhibit heterogeneous protein patterns, which cells control. Strong patterns, like the polarity spot in budding yeast, can be described as surface condensates, formed by physical interactions between constituents. However, it is unclear how these interactions affect the material exchange with the bulk. To study this, we analyze a thermodynamically consistent model, which reveals that passive exchange generally accelerates the coarsening of surface condensates. Active exchange can further accelerate coarsening, although it can also fully arrest it and induce complex patterns involving various length scales. We reveal how these behaviors are related to non-local transport via diffusion through the bulk, rationalizing the various scaling laws we observe and allowing us to interpret biologically relevant scenarios.
\end{abstract}
\maketitle

\tableofcontents

\section{Introduction}

Biological cells are enclosed by lipid membranes, which are crucial for interacting with the surrounding.
These membranes are complex, and cells control their composition in space and time.
In part, they achieve this by exploiting phase separation to form condensates embedded in the membrane~\cite {Banjade2014, McAffee_2022, Litschel2024, Kamatar2024}.
One example is the polarity spot of budding yeast, which is a dense accumulation of multiple proteins at one region of the membrane, forming in preparation of cell division~\cite{Chiou_2017}.
In this case, the proteins exchange with the bulk cytosol is apparently controlled by cells to ensure faithful division~\cite{Gulli2000}.  %
However, the general principles of controlling condensates in membranes by material exchange remain elusive.

Models of pattern formation on membranes have been discussed in various context ranging from cluster descriptions~\cite{Turner2005} to spatially resolved models.
Prominent examples of the latter are reaction-diffusion models~\cite{Lang2022, John_2005_PRL, John_2005_PB, Gessele2020, Brauns_2021}, which explain the emergence of patterns, but cannot describe physical interactions required for phase separation. %
Phase separation has been included in various thermodynamic models, e.g., in the specific case of surface wetting~\cite{Zhao2021, Zhao2024,Zhao_2025} and in more general theories that prescribe fixed exchange rates~\cite{Foret2005, Caballero_2023}.
However, it is unclear how thermodynamic constrains from physical interactions impact material exchange and the resulting patterns.

\begin{figure}[t]
  \centering
   {\includegraphics[width= 1\columnwidth]{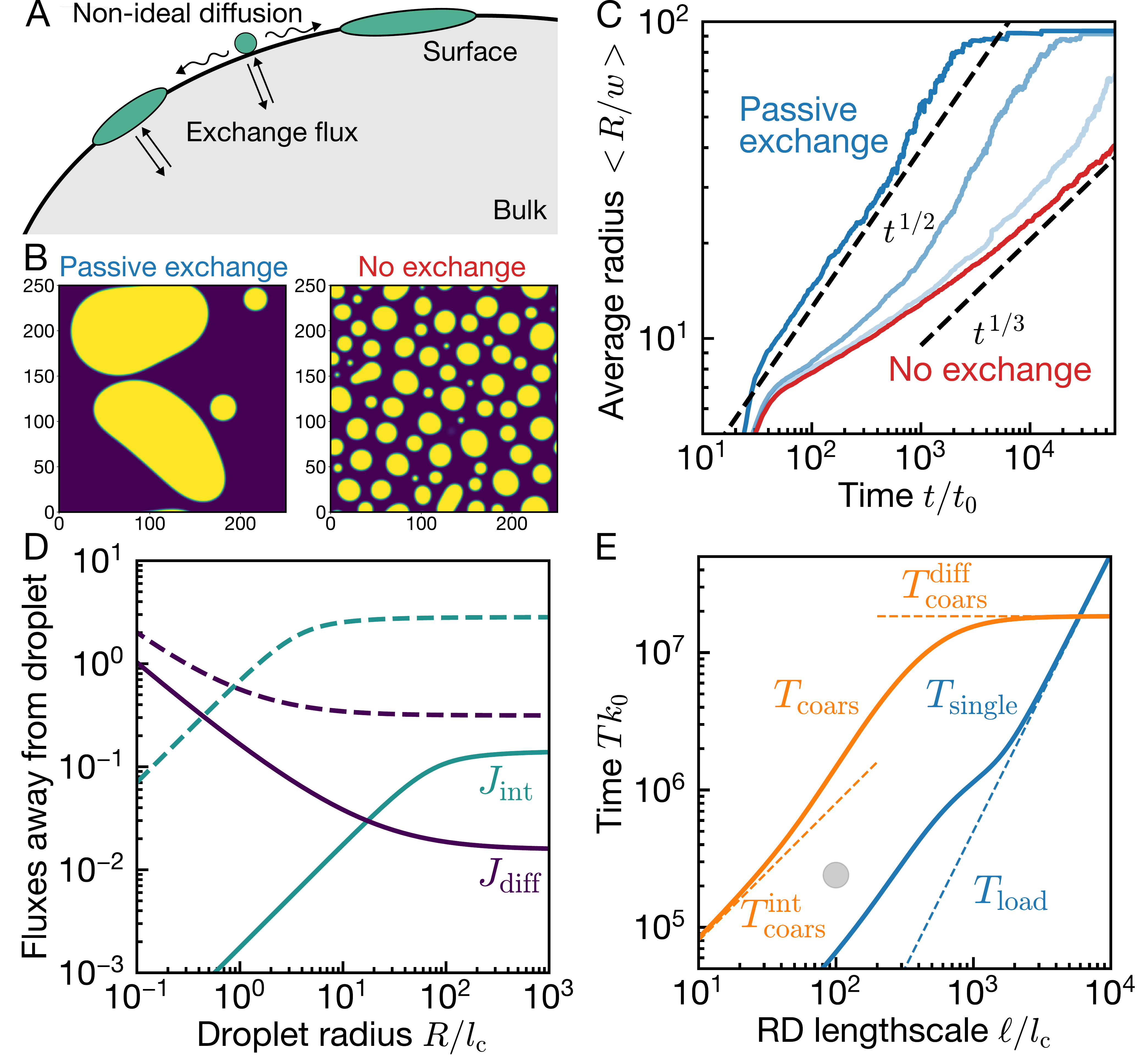}}
   
  \caption{
  	\textbf{Passive exchange with bulk accelerates coarsening in a surface.}
	(A) Schematic of our model describing interacting solutes that phase separate in a 2D surface and exchanges with a 3D bulk. 
	(B) Snapshots of numerical simulations with and without passive exchange at the same time $t=10^3\,t_0$ for $\PhiTot=0.5$, $\chi =3.5$, $\kappa =2 \, \kBT w^2 / \nus$, $\omega_\mathrm{b} -\omega_\mathrm{s}= 6.5$, $\gamma A = 100$, $\ka =0$, $w = (\nus \kappa/ \kBT)^{1/2}$, and $t_0 = w^2/(\kBT \Lambda)$.
	(C) Mean droplet radius~$R$ as a function of $t$ for $\kp=0$ (red line) and $\kp t_0 = 10^{-3}, 10^{-2}, 10^{-1}$ (lighter to darker blue lines) for the same parameters as in (A).
	(D) Fluxes from a droplet toward the dilute phase ($\Jdiff$) and the bulk ($\Jint$) as a function of radius $R$ for reaction-diffusion lengths $\ell_\mathrm{in} = \ell_\mathrm{out}=\ell=40 \, \ellc$ (solid lines) and $\ell=2 \, \ellc$ (dashed lines) for $\phi_\mathrm{out}^{(0)} = 0.1$, $\phi_\mathrm{in}^{(0)} = 0.9$, $k_0=D/\ellc^2$ with $D=D_\mathrm{in} = D_\mathrm{out}$.
	(E) Time scale of growing a single droplet ($T_\mathrm{single}$, blue line) and coarsening of two droplets ($T_\mathrm{coars}$, orange line) as a function of~$\ell$.
	Limiting cases ($T_\mathrm{coars}^{\mathrm{diff}}$, $T_\mathrm{coars}^{\mathrm{int}}$, and $T_\mathrm{load}$) are discussed in  main text. 
	Gray disk indicates estimates for polarity spots in budding yeast (\appref{app:timescales_estimation}).
	Parameters are $R_* = 10^2 \, \ellc$, $L= 10^3 \, \ellc$, $\PhiTot=0.5$, $\eta = 100$, $\phi_\mathrm{dil} =0.101$, and given in (D).
}
\label{fig:passive_coarsening}
\end{figure}

In this letter, we develop a minimal, thermodynamically consistent model of phase separation in surfaces including passive and active material exchange with the bulk (\figref{fig:passive_coarsening}A). 
We find that passive exchange generally accelerates coarsening, whereas activity can either accelerate or suppress it, depending on details.
 We characterize the state of the surface by the area fraction field $\phis(\vect r,t)$ of solute molecules, so the solvent fraction is $1-\phis(\vect r,t)$.
In contrast, the state of the bulk is described by the mean solute fraction $\phib(t)$ 
since we assume fast diffusion in the bulk. 
The associated dynamics are governed by
\begin{subequations} \label{eq:dynamics}
\begin{align}
	\partial_t \phis(\vect r, t) & =  \Lambda \nabla^2 \mus(\vect r, t) + s(\vect r, t)
	\label{eq:dynamics_surface}
\\
	\partial_t \phib(t) & = - \gamma \int_As(\vect r, t) \, \diff A
	\label{eq:dynamics_bulk}
	\;,
\end{align}
\end{subequations}
where $\Lambda$ is the solute mobility on the surface.
\Eqref{eq:dynamics_bulk} describes the effect of the exchange flux~$s$ integrated over the entire surface~$A$ and the parameter $\gamma = \nub/(\nus V)$ captures the relative size of the surface to the bulk with volume~$V$.
Here, $\nub$ denotes the molecular volume in the bulk, wheres $\nus$ is the molecular area in the surface. %
Finally, we express the exchange flux $s$ as~\cite {Zwicker_2025, Zwicker_2022}
\begin{equation} \label{eq:active_binding_flux}
	s = 2 k_\mathrm{p}  \sinh \Bigl( \frac{ \mub - \mus}{2 \kBT} \Bigr) 
	+ 2k_\mathrm{a}  \sinh \Bigl( \frac{ \mub - \mus -\Delta \mu}{2 \kBT} \Bigr)
	\;,
\end{equation}
where $\kBT$ sets the energy scale.
The first term describes passive exchange proportional to the rate  $k_\mathrm{p}>0$.
This exchange obeys detailed balance and is driven by difference between the bulk chemical potential $\mub(t)$ and the surface chemical potential $\mus(\vect r, t)$, which quantify the increase in free energy when solvent is replaced by a solute particle  in the bulk or surface, respectively.
In contrast, the second term in \Eqref{eq:active_binding_flux} describes an active exchange proportional to the rate $k_\mathrm{a}>0$, where the chemical potential difference is biased by the external energy input $\Delta \mu$ provided by a fuel~\cite {Kirschbaum_2021,Zwicker_2025}. 
The two chemical potentials, $\mus = \nus \delta F[\phis,\phib]/\delta \phis$ and $\mub =(\nub/\Vb) \partial F[\phis,\phib]/\partial \phib$, follow from the free energy~$F$,
\begin{equation} \label{eq:free_energy}
	F[\phis,\phib] = \int_A \Bigl[ f_\mathrm{s}(\phis) + \frac{\kappa}{2} | \nabla \phis|^2 \Bigr] \diff A + \Vb f_\mathrm{b} (\phib) \;,
\end{equation}
where the local free energy densities $f_\mathrm{s}$ and $f_\mathrm{b}$ govern the behaviors of homogeneous mixtures in the surface and bulk, respectively.
Analyzing equilibrium states reveals that interesting behavior requires phase separation in the surface~\cite{Rossetto_2025}, which we describe by a Flory--Huggins energy, $f_\mathrm{s} = \kBT \nus^{-1} [\chi \phis (1-\phis) + \phis \log(\phis) + (1-\phis) \log(1-\phis) + \omega_\mathrm{s} \phis]$.
Here, $\chi$ accounts for interactions driving phase separation, leading to the coexisting volume fractions $\phi_\mathrm{in}^{(0)}$ and $\phi_\mathrm{out}^{(0)}$~\cite{Qian2022}.
The gradient term proportional to $\kappa$ in \Eqref{eq:free_energy} penalizes concentration variations in the surface, leads to interfaces of finite width, and induces surface tension.
We assume that interactions do not play a significant role in the bulk, which we thus describe as an ideal mixture, $ f_\mathrm{b} = \kBT/\nub [ \phib \log(\phib) + \phib \omega_\mathrm{b}]$.
The energy difference $\omega_\mathrm{b} - \omega_\mathrm{s}$ controls the affinity of the molecules toward the surface, and
we neglect explicit coupling terms describing wetting since they typically just lead to parameter rescaling~\cite{Rossetto_2025}. 
Taken together, \Eqsref{eq:dynamics}--\eqref{eq:free_energy} describe thermodynamically consistent dynamics of interacting solute molecules that can form patterns in the surface and exchange with the homogeneous bulk.

\section{Results}

\subsection{Passive exchange leads to faster coarsening}

To understand the dynamics of phase separation in the surface, we first consider passive systems ($\ka=0$). %
In particular, we perform numerical simulations of \Eqsref{eq:dynamics}--\eqref{eq:free_energy} for a two dimensional surface with periodic boundary conditions~\cite {Zwicker_2020}, which leads to droplets of various sizes (\figref{fig:passive_coarsening}B). 
We quantify the dynamics of these droplets by their average radius as a function time (\figref{fig:passive_coarsening}C).
Without exchange ($\kp=0$, red line), \Eqref{eq:dynamics_surface} reduces to regular Cahn--Hilliard dynamics~\cite {Cahn_1958}, explaining the observed Lifshiz--Slyozov coarsening of $R \sim t^{1/3}$~\cite {Lifshitz_1961, Marqusee_1984}.
Enabling passive exchange ($\kp>0$, blue lines) accelerates coarsening, which is expected since the exchange provides an additional transport channel between droplets.
A linear stability analysis of the equations supports this conclusion since it shows that $\kp$ increases the growth rate of perturbation modes with long wavelengths (\appref{app:LSA}).
Beside the quantitative acceleration of the coarsening, we also observe a transition to a qualitatively different coarsening regime, where the coarsening law approaches $R \sim t^{1/2}$ for sufficiently large droplets.

\subsection{Diffusive efflux and internal unbinding control coarsening}
\label{sec:passive_effective_model}
To study why passive exchange can alter coarsening qualitatively, we next evaluate the fluxes between droplets that formed in the surface.
Focusing on an individual droplet of radius $R$, the equilibrium fractions $\phi_n^\mathrm{eq}= \phi_n^{(0)}( 1+ \ellc/R)$ for $n \in \{ \mathrm{in},\mathrm{out} \}$ are slightly elevated by surface tension effects, which is quantified by  the capillary length $\ellc$~\cite{Zwicker_2025}. 
We approximate the dynamics inside and outside the droplet by expanding \Eqsref{eq:dynamics} around the coexisting fractions $\phi^{(0)}_\mathrm{in}$ and $\phi^{(0)}_\mathrm{out}$, %
\begin{equation} \label{eq:effective_droplet}
	\partial_t \phis_{n} \simeq D_{n} \nabla^2 \phis_{n} - k_{n} (\phis_n - \phi_{n}^{(0)})
	\;;
\end{equation}
for $n \in \{ \mathrm{in},\mathrm{out} \}$; 
see \appref{app:eff_drop}
Combining the resulting diffusivities $D_n \simeq \Lambda \nus f_\mathrm{s}''(\phi_{n}^{(0)})$ and the rate constants $k_n = - s'(\phi_n^{(0)})$, we define 
the reaction-diffusion length scales $\ell_n = (D_n/k_n)^{1/2}$, which quantify how far solute molecules diffuse in the surface before  unbinding.
Mass balance implies the droplet growth rate~\cite{Zwicker_2025}
\begin{equation} \label{eq:d_radius}
	\frac{\diff R}{\diff t} = - \frac{\Jdiff +\Jint}{2 \pi R \bigl(\phi^{(0)}_\mathrm{in} - \phi^{(0)}_\mathrm{out}\bigr)}
	\;,
\end{equation} 
where 
$\Jdiff = -2 \pi R D_\mathrm{out}  \phis_\mathrm{out}'(R)$ 
denotes the diffusive flux from the interface to the surrounding dilute phase, whereas $\Jint$ captures the internal unbinding flux from the droplet to the bulk. 
In the quasi-stationary limit, $\Jint$ equals the the internal unbinding flux, implying $\Jint = 2 \pi k_\mathrm{in} \int_0^R r (\phi_{\mathrm{in}}(r) -  \phi_\mathrm{in}^{(0)}) \mathrm{d}r$ for a polar symmetric system.
We next show  that these two fluxes can lead to very different coarsening behavior.

To obtain an expression for the flux~$\Jdiff$ from the droplet interface toward the dilute phase, we solve \Eqref{eq:effective_droplet} at stationarity assuming polar symmetry.
Using the boundary condition $ \phis_{\mathrm{out}}(R) =  \phi_\mathrm{out}^\mathrm{eq}(R)$ at the interface and no-flux  conditions far away from the droplet, we find %
\begin{equation}
	\Jdiff =  2 \pi  R  \frac{D_\mathrm{out}}{\ell_\mathrm{out}}\frac{ K_1(R/\ell_\mathrm{out})}{K_0(R/\ell_\mathrm{out})} 	\Bigl(\phi_\mathrm{out}^\mathrm{eq}(R) - \phi_\mathrm{out}^{(0)}\Bigr)\;,
	\label{eqn:flux_diff}
\end{equation}
where $K_0$, $K_1$ are modified Bessel functions of the second kind.
If the exchange is slow ($ R \ll \ell_\mathrm{out}$), $\Jdiff$ behaves as a purely diffusive flux and $\Jint$ is negligible, leading to the expected Lifshitz--Slyozov scaling of $R \sim t^{1/3}$; see \appref{app:diff_int}.
In the converse limit ($R \gg \ell_\mathrm{out}$), fast exchange implies a homogeneous dilute phase, $\phi_\mathrm{out}(r)=\phi_\mathrm{out}^{(0)}$, so $\Jdiff$ becomes proportional to the size of the droplet's interface, implying $R \sim t^{1/2}$~\cite {Wagner1961, Lee_2021}; see \appref{app:diff_int}.

The internal unbinding flux $\Jint$ follows from an analogous evaluation of the profile $\phis_{\mathrm{in}}(r)$  with boundary conditions $\phis_{\mathrm{in}}(R) = \phi_\mathrm{in}^\mathrm{eq}(R)$ and $\phis_{\mathrm{in}}'(r=0)=0$, 
\begin{equation} 
	\Jint = 2 \pi R \frac{D_\mathrm{in}}{\ell_\mathrm{in}} \frac{I_1(R/\ell_\mathrm{in})}{I_0(R/\ell_\mathrm{in})} \Bigl(\phi^\mathrm{eq}_{\mathrm{in}}(R) - \phi^{(0)}_\mathrm{in}\Bigr) \;,
	\label{eqn:flux_int}
\end{equation}
where $I_0$, $I_1$ are modified Bessel functions of the first kind.
For slow exchange ($R \ll \ell_\mathrm{in}$), diffusion equilibrates the profile inside the droplet, $\phis_{\mathrm{in}}(r) \simeq \phi^\mathrm{eq}_{\mathrm{in}}(R) $, so $\Jint$ becomes proportional to the droplet's area.
Assuming that internal unbinding dominates transport ($\Jdiff \simeq 0$),
 this regimes leads to a linear scaling $R \sim t$.
Conversely, if $R \gg \ell_\mathrm{in}$, the material unbinds before being able to diffuse across the droplet, so material exchange is confined to a region close to the interface, implying $R \sim t^{1/2}$. %
These coarsening laws are indeed observed in effective numerical simulations (\appref{app:numerical_details}).

The effective fluxes allow us to interpret the numerically observed coarsening (\figref{fig:passive_coarsening}C).
When the exchange is fast in both phases  ($R \gg \ell_n$), the internal unbinding flux $\Jint$ behaves as an interface flux and dominates  transport (\figref{fig:passive_coarsening}D), leading to the scaling $R \sim t^{1/2}$ observed for the dark blue curve in \figref{fig:passive_coarsening}C.
Conversely, when the exchange is slow ($R \ll \ell_n $), diffusion dominates ($\Jdiff \gg \Jint$, \figref{fig:passive_coarsening}D), implying $R \sim t^{1/3}$.
For intermediate exchange kinetics (light blue curves in \figref{fig:passive_coarsening}C), we find that diffusion dominates early, when droplets are small and close to each other, whereas exchange dominates later for large droplets that are far apart.
In contrast, the ballistic regime ($R\sim t$), is inaccessible because it would require large $\Jint$ in the slow exchange regime ($R \ll \ell_n$) where $\Jdiff$ dominates instead; see \figref{fig:passive_coarsening}D.
Our analysis highlights that the two different fluxes can lead to qualitatively different coarsening regimes, which further depend on the relative size of droplets to the reaction-diffusion length scale.

\subsection{Coarsening with passive exchange cannot explain fast formation of polarity spot}
\label{sec:timescales}

We next use the effective fluxes to approximate the relevant timescales governing the dynamics of single droplets and droplet pairs, where we for simplicity consider $\ell_\mathrm{in} = \ell_\mathrm{out} = \ell$ and $D_\mathrm{in} = D_\mathrm{out} = D$. 
We start by estimating how long it takes to form one large, stationary droplet in a finite system~\cite{Zwicker_2015}.
We estimate the corresponding time scale $T_\mathrm{single}$ from a linear stability analysis around this stationary state (\appref{app:timescales_estimation}). 
This time scale describes the exchange of material between bulk and surface, which is slow for large $\ell$ (\figref{fig:passive_coarsening}E). 
In this limit, $T_\mathrm{single}$ is simply given by $T_\mathrm{load} \sim \ell^2/2D$, which describes homogeneous exchange. 
In contrast, for small $\ell$, a droplet in the surface exchanges material with the bulk only in a region close to its interface, leading to slower dynamics.
Our analysis shows that $T_\mathrm{single}$ grows with $\ell$, highlighting that passive exchange accelerates droplet formation.

For a passive system, multiple droplets always coarsen until one droplet remains.
We estimate the corresponding coarsening time~$T_\mathrm{coars}$ from the rate at which the sizes of droplets separated by a distance $L$ diverge (\appref{app:timescales_estimation}). 
\figref{fig:passive_coarsening}E shows that $T_\mathrm{coars}$ exceeds $T_\mathrm{single}$, essentially because material exchange between droplets is slower than  exchange with the bulk.
Moreover, the two time scales exhibit different scaling laws as a function of $\ell$:
For fast exchange ($R,L \gg \ell$), we find  (\appref{app:timescales_estimation})
\begin{equation} \label{eq:time_int}
	T_\mathrm{coars}^\mathrm{int} \simeq \frac{R_*^2 \ell ( \phi_\mathrm{in}^{(0)}-\phi_\mathrm{out}^{(0)})} {D \ellc( \phi_\mathrm{out}^{(0)}+\phi_\mathrm{in}^{(0)})} \;,
\end{equation}
which is independent of the separation $L$ since diffusive and internal unbinding fluxes are interface-limited.
In contrast, for slow exchange ($R,L \ll \ell$), the internal unbinding flux is area-limited and molecules diffusing between droplets rarely unbind from the surface,
leading to the timescale (\appref{app:timescales_estimation})
\begin{equation} \label{eq:time_area+diff}
	T_\mathrm{coars}^\mathrm{diff} \simeq \frac{R_*^3 \log (L/R_*)( \phi_\mathrm{in}^{(0)}-\phi_\mathrm{out}^{(0)})}{D \ellc \phi_\mathrm{out}^{(0)}}
	\;,
\end{equation}
which is independent of $\ell$, because the dynamics are essentially confined to the surface.

We next compare the estimated time scales to experimental measurements of polarity spots in budding yeast~\cite{Chiou_2017}.
We estimate the diffusivity $D \sim \SI{0.2}{\micro \meter^2} / \mathrm{s}$ and the exchange rate $k \sim \SI{0.2}{s^{-1}}$ using measured values for the key protein Cdc42~\cite{Rutkowski_2024}.
Additionally assuming a capillary length of $\ellc \sim  \SI{10}{\nano \meter}$ based on molecular sizes, and considering two polarity spots of size $R_* \sim \SI{1}{\micro \meter}$ at distance $L \sim \SI{10}{\micro \meter}$, we find  $T_\mathrm{single} \sim  \SI{30}{\second}$ and $T_\mathrm{coars} \sim \SI{7e2}{\second} $.
Since polarity spots form within about $2$ minutes, this suggests that passive exchange is consistent with their formation, but it fails to account for sufficiently fast resolution of two spots.
Instead, we would expect that evolution devised a process that resolves multiple spots significantly more quickly to ensure robust formation of the polarity spot.
This suggests that the experimentally observed activity~\cite {Chiou_2017} is crucial.

\subsection{Homogeneous activity leads to futile cycles}

To study active unbinding from the surface ($\ka \neq 0$), we first consider the simple case where the exchange is independent of position ($\kp, \ka, \Delta\mu$ constant), the behavior is qualitatively similar to the passive case. 
To see this, we linearize the exchange flux given by \Eqref{eq:active_binding_flux} using $\sinh(x) \approx x$, which implies
$s_\mathrm{p} = - s_\mathrm{a} \approx \Delta \mu \kp \ka /[\kBT (\kp + \ka)]$ at stationarity.
The  system thus never reaches equilibrium when $\ka$ and $\Delta\mu$ are finite.
Specifically, for $\Delta \mu >0$, material is removed from the surface using the external energy input, whereas it rebinds through the passive exchange flux.
However, if the rates $\kp$ and $\ka$ are constants, these fluxes are balanced everywhere, leading to futile cycles, which do not alter the dynamics~\cite{Kirschbaum_2021}.

To find interesting behavior, we need to break the futile cycles so that binding toward the surface at one place is balanced by unbinding at another place.
To do this, we consider constant $\kp$ combined with composition-dependent active unbinding, 
\begin{align}
	\kp &= k
&
	\ka &=k  \Bigl[ \frac{1+\alpha}{2} \phis + \frac{1-\alpha}{2} (1 - \phis) \Bigr]
	\;,
\end{align}
motivated by enzymes segregating into droplets~\cite{Kirschbaum_2021}.
For simplicity, we here assume that passive and active exchange occur with similar rate~$k$, but the active flux is modified by the asymmetry parameter $\alpha\in[-1,1]$.
For $\alpha>0$ and $\Delta\mu>0$, active unbinding happens predominately inside droplets, where $\phi$ is large.
Conversely, for $\alpha < 0$ (but still $\Delta\mu>0$), the active flux is biased toward dilute regions, and $\alpha=0$ corresponds to homogeneous active exchange, which exhibits passive dynamics.
We can thus use $\alpha$ to control the direction of circular fluxes between dense and dilute surface regions and the bulk.

\begin{figure}
  \centering
  {\includegraphics[width= 1\columnwidth]{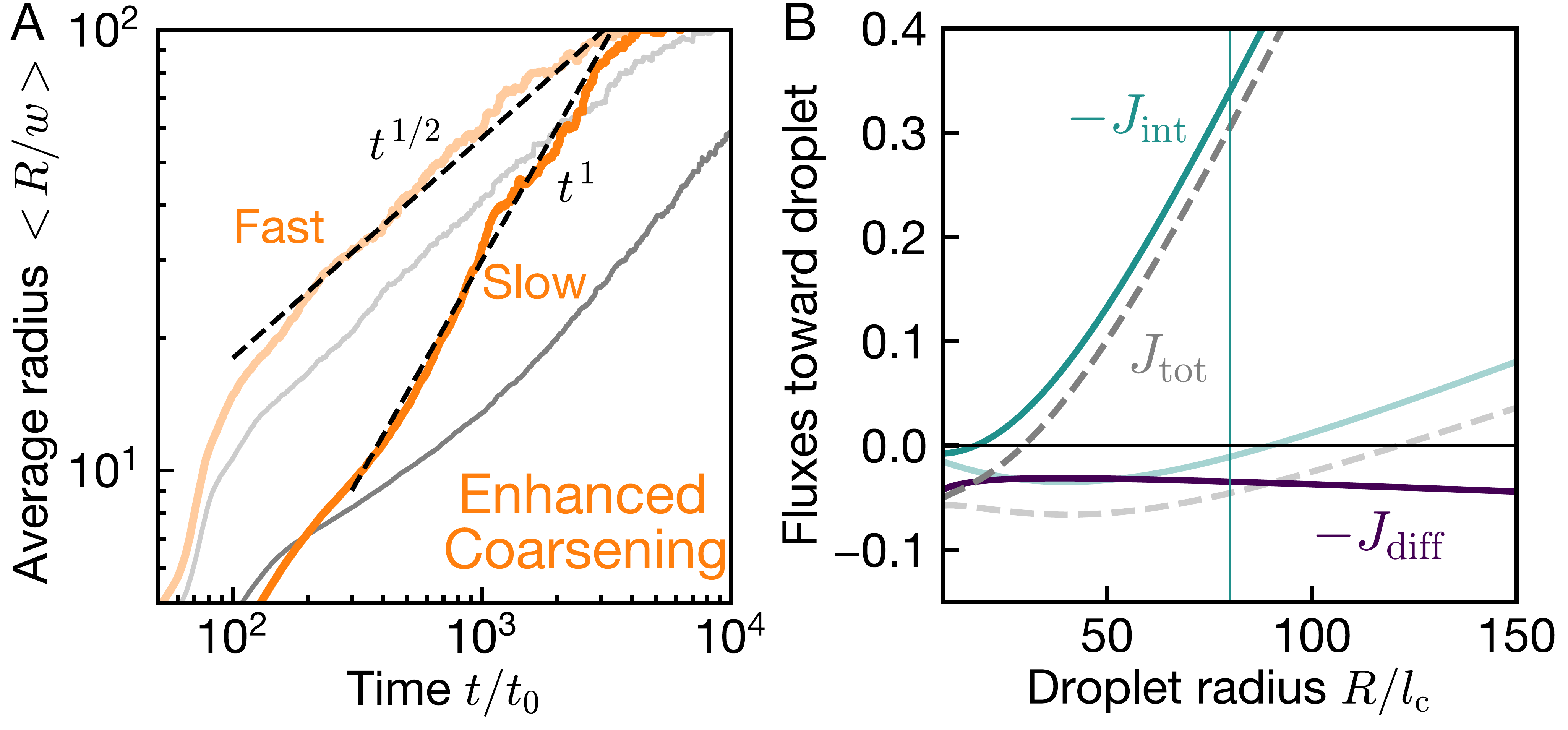}}
 
  \caption{
	\textbf{Active unbinding from dilute regions accelerates coarsening.}
	(A) Mean droplet radius $R$ as a function of time $t$ for $\alpha=-0.5$ (active system, orange line) and $\alpha=0$ (passive system, gray line) as well as $k=10^{-1} /t_0$ (light colors) and $k=10^{-2}/t_0$ (dark colors).
	Parameters are $\Delta \mu=2\,\kBT$ and given in \figref{fig:passive_coarsening}(B--C).
	(B) Diffusive efflux~$J_\mathrm{diff}$ and internal unbinding flux $J_\mathrm{int}$ as a function of $R$ for $\Delta\phi_\mathrm{in} = 0.01$ (light colors) and $\Delta\phi_\mathrm{in} = 0.05$ (dark colors).
	Grey dashed line corresponds to $J_\mathrm{tot} = \Jdiff + \Jint$.
	Vertical line indicates estimated transition between area- and interface-limited flux  $\Jint$ (\appref{app:diff_int}).
	Model parameters are $\Delta\phi_\mathrm{out} = -10^{-3}$, $\ell = 40 \,\ellc$, and given in \figref{fig:passive_coarsening}(D--E).
}
\label{fig:enhanced_coarsening}
\end{figure}

\subsection{Active unbinding from dilute regions accelerates coarsening} \label{sec:active_enhanced}

We first focus on the case where activity promotes unbinding from dilute regions ($\alpha<0$).
\figref{fig:enhanced_coarsening}A shows that such activity (orange lines) accelerates coarsening compared to the passive case (gray lines).
The data suggests that not only is the pre-factor increased, but also that the coarsening exhibits a different power-law, where the average droplet radius scales linearly with time for slower exchange.
To understand this behavior, we again study the fluxes $\Jdiff$ and $\Jint$ affecting a droplet of radius $R$.
In contrast to the passive case discussed above, the exchange flux now does not vanish at phase equilibrium ($\phi_\mathrm{in} = \phi^{(0)}_\mathrm{in}$ and $\phi_\mathrm{out} = \phi^{(0)}_\mathrm{out}$).
The difference between phase equilibrium and exchange equilibrium is quantified by fractions $\Delta\phi_n$, 
which imply the following reaction-diffusion equations in the phases $n=\text{in}, \text{out}$~\cite{Weber_2019,Zwicker_2022},
\begin{equation} \label{eq:effective_droplet_active}
	\partial_t \phis_{n} \simeq D_{n} \nabla^2 \phis_{n} - k_{n} (\phis_n - \phi_{n}^{(0)} - \Delta\phi_n) \;,
\end{equation}
where $k_n=-s'(\phizn)$ and $\Delta\phi_n = s(\phi_n^{(0)}) /k_n$. 
For passive systems ($\Delta\phi_n=0$), this equation is identical to \Eqref{eq:effective_droplet}.
Consequently, the associated fluxes $\Jdiff$ and $\Jint$ can be obtained via replacing $\phi^{(0)}_n$ by $\phi^{(0)}_n + \Delta\phi_n$ in \Eqsref{eqn:flux_diff} and \eqref{eqn:flux_int}, respectively (\appref{app:eff_drop}).
For $\alpha<0$, we find $\Delta\phi_\mathrm{out}<0$ and $\Delta\phi_\mathrm{in}>0$, which enables systems where $\Jint$ dominates for small droplets ($R\ll \ell_\mathrm{in}$, \figref{fig:enhanced_coarsening}B), which is impossible in passive systems (\figref{fig:passive_coarsening}D) for $\ell_\mathrm{in}= \ell_\mathrm{out}$. 
In this case, we predict ballistic growth ($R \sim t$), consistent with our data (\figref{fig:enhanced_coarsening}A), and even super-linear scalings become accessible for strong activity (\figref{fig:app_1}C).
However, our theory predicts a transition to $R\sim t^{1/2}$ when droplets are sufficiently large, so that $\Jint$ eventually becomes interface-limited (\appref{app:diff_int}).
For faster exchange rates, this transition takes place at smaller droplets, implying that $R\sim t^{1/2}$ is observed earlier, consistent with data (\figref{fig:enhanced_coarsening}A).

Overall, our numerical simulations indicate that active unbinding in dilute regions accelerates coarsening. 
Qualitatively, this is because active unbinding effectively undersaturates the dilute region ($\Delta\phi_\mathrm{out} < 0$), while adding material to dense regions ($\Delta\phi_\mathrm{in} > 0$).
The resulting circular fluxes enhance material transport from small to large droplets, thus accelerating coarsening. 
To investigate this acceleration quantitatively, we again estimate the time $T_\mathrm{coars}$ needed to resolve two spots separated by~$L$.
Using the same technique as above and considering parameters compatible with polarity formation in budding yeast, we find that a relatively small energy input ($\Delta \mu =2\,\kBT$) can lead to significant acceleration ($T_\mathrm{coars}^\mathrm{act} \sim \SI{e2}{\second}$, \appref{app:act_coars_time}).
This suggests that budding yeast uses active processes to promote unbinding from dilute regions to ensure the formation of a single polarity spot quickly and robustly.

\subsection{Active unbinding from dense regions can arrest coarsening}

\begin{figure}
  \centering
  \includegraphics[width=\columnwidth]{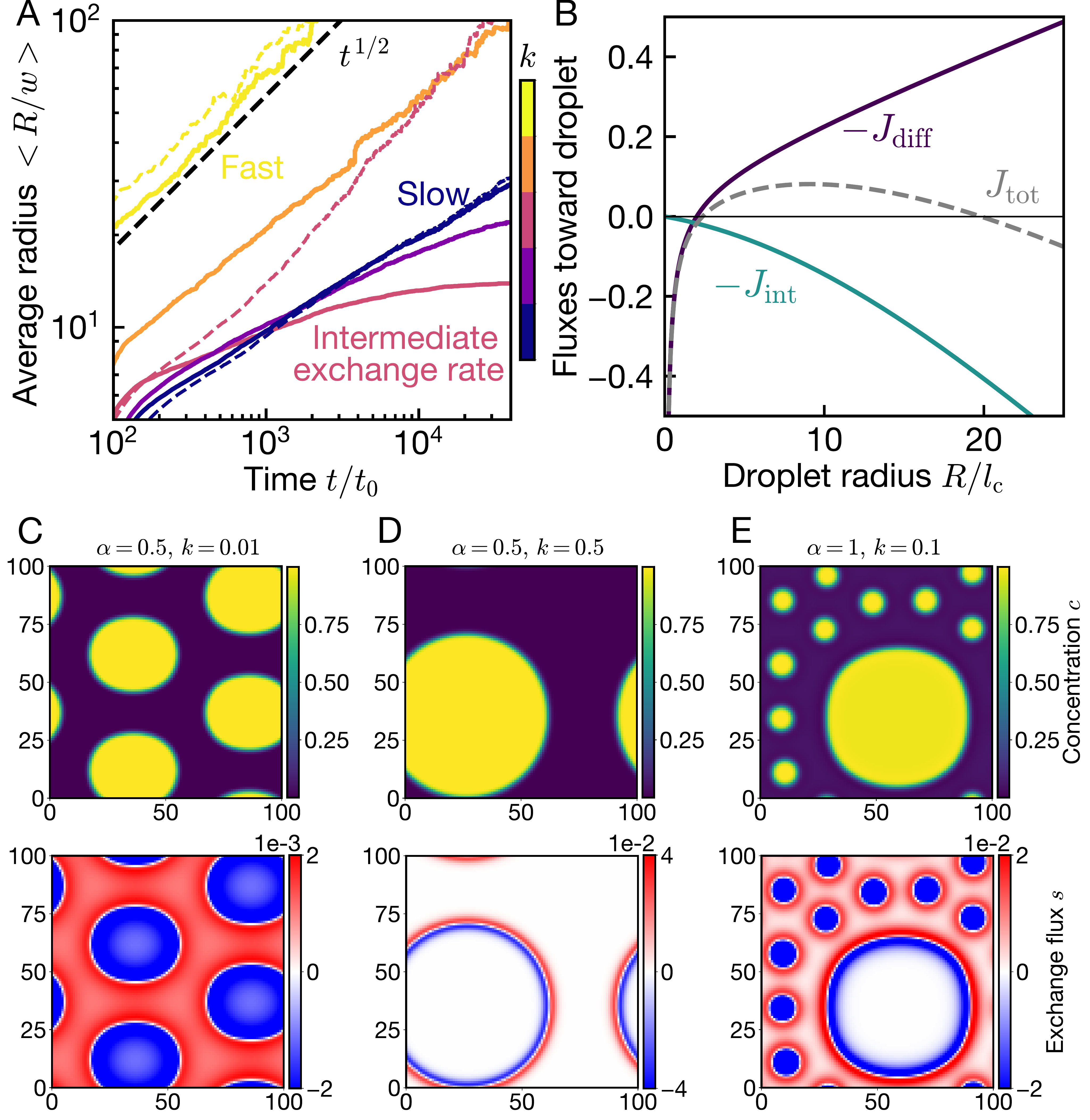}
  \caption{
	\textbf{Active unbinding from dense regions can arrest coarsening.}
	(A) Mean droplet radius $R$ as a function of time $t$ for $\alpha=0.5$, $\Delta \mu=2 \,\kBT$, and $ k t_0= 5 \times 10^{-1}, 10^{-1}, 10^{-2},10^{-3}, 10^{-4}$ (fast to slow).
	Dashed curves corresponds to passive limits ($\alpha = 0$).
	Remaining parameters are given in \figref{fig:passive_coarsening}(B--C).
	(B) Diffusive efflux and internal unbinding fluxes for $\Delta\phi_\mathrm{in} = -0.1$, $\Delta\phi_\mathrm{out} = 0.05$, $\ell=10\, \ellc$. Remaining parameters and units are given in \figref{fig:passive_coarsening}(D--E). The grey dashed line corresponds to $J_\mathrm{tot} = \Jdiff + \Jint$.
	(C--E)
	Top panels: Stationary concentration field $c(\vect r, t)$ (top panels) and corresponding exchange fluxes (bottom panels) for various parameters:
	(C) $k = 10^{-2} /t_0$ (corresponding to pink line in panel~A),
	(D) $k = 5 \times 10^{-1} /t_0$ (yellow line in panel~A), and
	(E) $\alpha = 1$ with $\chi=3$, $\PhiTot=0.6$, and $k=0.1$.
	Remaining parameters are the same as in panel A.
}
\label{fig:active_binding}
\end{figure}

In the opposing case where activity promotes unbinding from dense regions ($\alpha>0$), coarsening is slowed (\figref{fig:active_binding}A).
Interestingly, the slow-down is strongest for intermediate exchange rates (pink line), where it seems to halt completely.
To understand this slow-down, we again study the total flux toward the droplet, $J_\mathrm{tot} = -\Jint - \Jdiff$, which governs droplet growth; see \Eqref{eq:d_radius}.
For $\alpha>0$, we find $\Delta \phi_\mathrm{out}>0$ and $\Delta \phi_\mathrm{in}<0$ (\appref{app:eff_drop}), which implies that $J_\mathrm{tot}$ exhibits two roots (\figref{fig:active_binding}B). 
While the first root corresponds to the critical radius below which droplets shrink, the second root is a stable state indicating that droplets of that size are stable.
This situation is akin to externally maintained droplets where chemical conversion of material drives similar fluxes, leading to arrested coarsening~\cite {Weber_2019, Kirschbaum_2021, Zwicker_2025}.
This similarity is also supported by linear stability analysis (\appref{app:LSA}), which reveals a band of unstable modes.
Taken together, material exchange between surface and bulk plays a similar role to the previously described reactions in bulk.
Intuitively, the arrested coarsening is caused by active unbinding from the dense region ($\Delta \phi_\mathrm{in}<0$), and rebinding at the dilute region ($\Delta \phi_\mathrm{out} >0$). 
The resulting fluxes transport material from large to small droplets, thus opposing coarsening and supporting regular hexagonal patterns (\figref{fig:active_binding}C).  
From analyzing chemically active droplets, we expect that smaller reaction rates (i.e., exchange rates in our case) lead to patterns with larger length scale~\cite{Glotzer_1994,Christensen_1996,Muratov_2002,Zwicker_2015,Koestler_2025}.
\figref{fig:active_binding}A indeed suggests that slower exchange affects coarsening less (blue line), and we predict that the average droplet radius reaches a stable stationary state eventually.
Taken together, we find that size-control of droplets eventually limits coarsening. 

We have shown that larger exchange rates arrest droplets at smaller sizes, implying a stronger suppression of coarsening.
While this is confirmed by \figref{fig:active_binding}A, very large exchange rates surprisingly lead to fast coarsening again,
which hardly differs from the passive case (yellow line in \figref{fig:active_binding}A). 
In these cases, we do not observe any size-control, and the system instead coarsens toward a single droplet (\figref{fig:active_binding}D).
The droplet sizes evolve as $R(t) \sim t^{1/2}$, consistent with passive exchange (\figref{fig:passive_coarsening}C).
For the large rates considered here, the exchange becomes interface-limited (\figref{fig:active_binding}D), implying a correction to the interface behavior, whereas phase coexistence behaves effectively as a passive system~\cite{Qiang2025,Cho_2025}.
Taken together, this suggest that strong reactions merely affect the interface, but the large-scale behavior is essentially unchanged.

In the case where activity promotes unbinding from dense regions, faster exchange kinetics (large $k$) generally accelerate coarsening when size-control is not yet relevant (early times in \figref{fig:active_binding}A), similar to passive exchange. 
However, in some cases active unbinding from dense regions introduces fluxes between droplets that can limit their sizes.
This size-control sets in earlier for larger exchange rates $k$~\cite{Koestler_2025}, explaining the observed earlier slow-down of coarsening (e.g., pink curve in \figref{fig:active_binding}A).
However, for even larger $k$, we do not observe any size-control, and instead the system coarsen indefinitely.
This distinction between size-control and coarsening suggests that there is a critical exchange rate, although  %
 we also observe multi-stability, where the behavior depends on initial conditions (\figref{fig:app_multi}).
Interestingly, we sometimes even observe coexistence between multiple droplets of equal size and a single large droplet (\figref{fig:active_binding}E), which we interpret as coexistence between a patterned phase and a dense macrophase.
In that case, we observe continuous coarsening of the macrophase, whereas the patterned phase plays the role of the dilute phase.

\section{Discussion}

In summary, we showed that passive exchange accelerates coarsening (\figref{fig:passive_coarsening}), whereas active exchange either accelerates (\figref{fig:enhanced_coarsening}) or suppresses coarsening (\figref{fig:active_binding}).
The precise coarsening dynamics depend on how droplets exchange material, which either happens by diffusion along the surface or via diffusion through the bulk upon unbinding.
The exchange dynamics determine whether these fluxes are limited to regions around the interface or not, which affects the coarsening dynamics qualitatively.
The suppressed coarsening is akin to chemically active droplets~\cite{Weber_2019,Zwicker_2025,Zwicker_2022}, which also exhibit multi-stability~\cite{Bauermann_2025,Koestler_2025}.
Beyond these similarities, we additionally observe the coexistence of many small droplets with one large homogeneous region.
Interestingly, similar patterns have been observed in phase-field crystals~\cite{Thiele2013} and systems with non-local elasticity~\cite{Qiang2024}.
In analogy, the exchange via the bulk might thus play the rolf of a non-local interaction in our system, suggesting that non-local interactions are key to controlling various patterns~\cite{Seul1995}. %

Our results have implications for membrane patterns in biology:
For instance, we propose that active exchange is necessary to quickly and robustly form only one polarity spot in budding yeast.
However, our general theory applies to many pattern forming systems, including PAR protein patterns in \textit{C. elegans}~\cite{Lang2017}, post-synaptic densities in neurons~\cite{Zeng2016}, LAT receptor clusters in bacteria~\cite{Benn_2021}, Min protein patterns~\cite{Loose2011,Brauns_2021,Wurthner2022}, and Noc proteins \textit{in vitro}~\cite{Babl2022}.
To describe these systems in detail, our model likely needs to be extended, e.g., by including dimerization~\cite{Bland_2024, Rossetto_2025}, advection~\cite{Goehring2011}, exchange with linear compartments~\cite{Ernst2025,Morin2022}, chemical reactions, membrane interactions~\cite{Winter2025}, and fluctuations.

\textit{Acknowledgements---}%
We thank Nynke Hettema, Stephan Köstler, Guido Kusters, Liedewij Laan, and Gerrit Wellecke for helpful discussions. 
We gratefully acknowledge funding from the Max Planck Society and the European Union (ERC, EmulSim, 101044662).

\bibliographystyle{apsrev4-1} 
\bibliography{bibliography}

\onecolumngrid
\newpage

\let\section\oldsection
\let\subsection\oldsubsection

\renewcommand{\thefigure}{S\arabic{figure}}
\setcounter{figure}{0}

\renewcommand{\theequation}{S\arabic{equation}}
\setcounter{equation}{0}

\begin{center}
  {\large\bfseries Supplementary Information\\[1em]}
\end{center}

\section{Thin interface approximation} \label{app:eff_drop}
We derive the thin-interface approximation by considering an individual droplet of radius $R$.
We can then approximate the thermodynamically consistent dynamics given by \Eqsref{eq:dynamics} in the main text inside and outside the droplet by perturbing the full equations around the coexistence fractions, using $\phi_n \approx \phi^{(0)}_n + \delta\phi_n$ for $n=\text{in}, \text{out}$.
We further assume that the overall binding equilibrium between surface and bulk is reached faster than time scales relevant for coarsening, implying that dynamics in the surface leave the bulk unchanged, $\delta \psi \sim 0$.
We thus obtain
\begin{equation}
\partial_t \delta \phi_n = \Lambda \nus f_\mathrm{s} ''(\phizn) \nabla^2 \delta \phi_n - \Lambda \kappa \nabla^4 \delta \phi_n + s(\phizn)+ s'(\phizn) \delta \phi_n \;,
\end{equation}
for $n = \mathrm{in},\mathrm{out}$.
Assuming that concentration variations are small, we neglect the fourth-order derivative,
\begin{equation}
	\partial_t \delta \phi_n = D_n\nabla^2 \delta \phi_n -k_n\delta \phi_n + s(\phizn) \;,
\end{equation}
where we also defined the diffusion constants $D_n = \Lambda \nus f_\mathrm{s}''(\phizn)$ and binding rates $k_n = -s'(\phizn)$. 
For passive systems, both the dense and dilute phases satisfy binding equilibrium~\cite {Rossetto_2025}, implying $ s(\phizn)=0$ and thus 
\begin{equation} \label{appA:eq:eff_dyn}
\partial_t \delta \phi_n = D_n\nabla^2 \delta \phi_n -k_n\delta \phi_n \;.
\end{equation}
In contrast, basal fluxes $ s(\phizn)$ are present in the active case.
To evaluate them, we consider the limit of small activity ($\Delta \mu \ll \kBT$), so the passive flux is close to equilibrium ($\mub - \mus \ll \kBT$) and we thus have
\begin{equation}
	s(\phizn) \simeq  k\Bigl( \frac{ \mub - \mus}{ \kBT} \Bigr)
	+ k \Bigl[ \frac{1 - \alpha}{2} + \alpha \phi \Bigr] \Bigl( \frac{ \mub - \mus -\Delta \mu}{ \kBT} \Bigr)
	\;.
\end{equation}
Further assuming strong phase separation $\phi_\mathrm{in}^{(0)} \sim 1$ and $\phi_\mathrm{out}^{(0)} \sim 0$ and absorbing a constant term in the bulk chemical potential $\tilde{\mu}_\mathrm{b} = \mub - \Delta \mu/(2 \kBT)$, we obtain the inside and outside exchange fluxes $s_\mathrm{in}$, $s_\mathrm{out}$
\begin{subequations} \label{eq:app:deltaphis}
\begin{align}
	s_\mathrm{in}(\phi_\mathrm{in}^{(0)}) & \simeq  k \Bigl( \frac{3 + \alpha}{2} \bigl[\tilde{\mu}_\mathrm{b} - \mus(\phi_\mathrm{in}^{(0)})\bigr] - \frac{\alpha \Delta \mu } {2 \kBT} \Bigr) \\
	s_\mathrm{out}(\phi_\mathrm{out}^{(0)}) & \simeq  k \Bigl( \frac{3 - \alpha}{2} \bigl[\tilde{\mu}_\mathrm{b} - \mus(\phi_\mathrm{out}^{(0)})\bigr]+  \frac{\alpha \Delta \mu } {2 \kBT} \Bigr)
	\;,
\end{align}
\end{subequations}
where $\mus (\phi_\mathrm{in}^{(0)}) = \mus (\phi_\mathrm{out}^{(0)})$ since we assume local equilibrium at the interface.
At stationarity, the two fluxes given by \Eqsref{eq:app:deltaphis} must have opposite signs to maintain coexisting phases.
Since $\alpha\in[-1,1]$, the first term in the fluxes have the same sign, implying that the second term must determine the sign.
In particular, we require $s (\phi_\mathrm{in}^{(0)}) >0$ for $\alpha<0$, and $s (\phi_\mathrm{in}^{(0)}) <0$ for $\alpha>0$, while the outside basal flux~$s_\mathrm{out}$ has opposing signs. 
Since $s'(\phizn) <0$ is needed for the stability of the dense and dilute states, we know that $k_n>0$, implying that the sign of the basal fluxes equals the sign of the corresponding fractions $\Delta \phi_n$.

\section{Linear stability analysis} \label{app:LSA}
\begin{figure}[t]
  \centering
    {\includegraphics[width= 1\columnwidth]{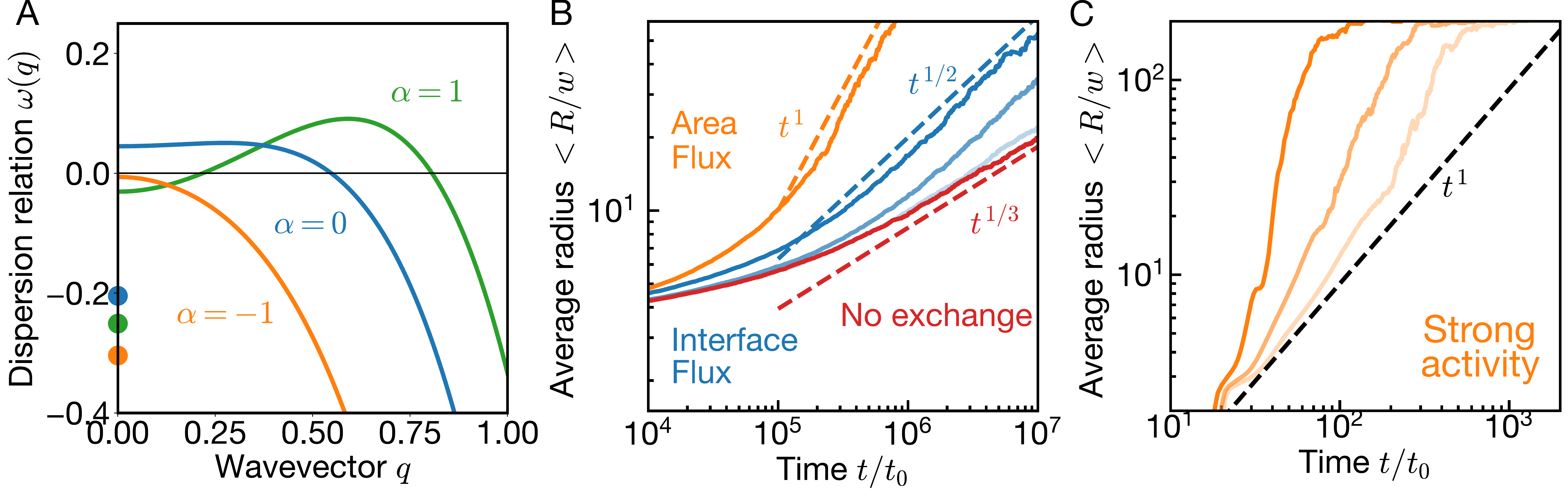}}
  \caption{ 
  (A) Dispersion relation from linear stability analysis for different choices of activity inhomogeneity coefficient $\alpha$. Parameters are $\chi = 2.7$, $\PhiTot = 0.6$, $k=0.1$. Remaining parameters are the same as \figref{fig:enhanced_coarsening}A.
  (B) Effective passive simulations showing scalings associated with different types of fluxes. 
  (C) Mean radius as a function of time for $\alpha =-0.5$, $\Delta \mu = 5 \, \kBT$, $\PhiTot = 1$. From lighter to darker orange the curves correspond to $k t_0=  5 \times 10^{-3}, 10^{-2},  3 \times 10^{-2} $. Remaining parameters are the same as \figref{fig:enhanced_coarsening}A in the main text.
}
\label{fig:app_1}
\end{figure}

To analyze the stability of a generic stationary state $(\phi^*(\vect r),\psi^*)$ of the thermodynamically consistent dynamics, we expand \Eqsref{eq:dynamics} of the main text by introducing the perturbations $\phi = \phi^* + \delta \phi(\vect r,t)$ and $\psi = \psi^* + \delta \psi(t)$.
The bulk perturbation $\delta \psi$ can be rewritten using mass conservation ($\bar{\phi} + \eta \psi = \Phi_\mathrm{tot}$), implying
\begin{equation}
	\label{eqn:bulk_perturbation}
\delta \psi(t) = \frac{1}{\eta A} \int_A \delta \phi(\vect r,t) \diff A = - \frac{\delta \hat{\phi}(0,t)}{\eta} \;,
\end{equation}
where we introduced the surface perturbation in Fourier space, $\delta \phi(\vect r,t) = \int \delta \hat{\phi} (\vect q,t) e^{i \vect q \vect r } \diff^d q$.
Perturbing the surface dynamics to linear order in $\delta\phi$, we find
\begin{equation}
 \partial_t \delta \phi = \Lambda \nuc f''_\mathrm{s} (\phi^*) \delta \phi - \Lambda \kappa \nabla^4 \delta \phi + s \;,
\end{equation}
where the exchange flux is given by
\begin{equation}
s \simeq \frac {k} {\kBT} \Bigl[ \bigl( \nub f''_\mathrm{b}(\psi^*) \delta \psi - \nuc f''_\mathrm{s}(\phi^*)\delta \phi + \kappa \nabla^2 \delta \phi \bigr) \bigl( 1 + C(\phi^*)\bigr) \\ + \alpha \delta \phi \bigl(\mu_\mathrm{b} (\psi^*) - \mu_\mathrm{c}(\phi^*) - \Delta \mu \bigr) \Bigr] \;,
\end{equation}
with $C(\phi) =  \frac{1+\alpha}{2} \phi + \frac{1-\alpha}{2} (1-\phi) $.
Transforming to Fourier space, we then find the dispersion relation
\begin{equation}
\tilde\omega(\vect q) = - \Lambda \kappa \vect q^4 - \Lambda  \vect q^2 \Bigl[ \nuc f''_\mathrm{c} (\phi^*) + \frac{k \kappa}{\kBT} \bigl( 1 + C(\phi^*)\bigr)  \Bigr] \\ + \frac{k} {\kBT} \Bigl[ \alpha \bigl(\mu_\mathrm{b} (\psi^*) - \mu_\mathrm{c}(\phi^*) - \Delta \mu \bigr) - \nuc f''_\mathrm{s}(\phi^*)  \bigl( 1 + C(\phi^*)\bigr)  \Bigr] \;,
\end{equation}
where we assumed $\vect q\neq 0$.
The special mode $\vect q=0$ requires a correction due to the bulk perturbation~$\delta\psi$, which can be written as the spatial average of the surface one; see \Eqref{eqn:bulk_perturbation}.
We thus find
\begin{equation}
	\omega(\vect q) = \tilde\omega(\vect q) -\begin{cases}
		\dfrac{k \nub}{ \eta \kBT} f_\mathrm{b}''(\psi^*) \bigl( 1 + C(\phi^*)\bigr)  & \vect q = \vect 0\\[10pt]
		0 & \vect q \neq \vect 0
	\end{cases}\;,
\end{equation}
leading to a discontinuous dispersion relation, which only depends on $q=|\vect q|$ because of isotropy.
We evaluate $\omega(q)$ for homogeneous stationary states obeying exchange flux balance.
For the passive case ($\alpha=0$), \figref{fig:app_1}A shows that $\omega(q) >0$ at $q \rightarrow 0$ and $\omega(q)<0$ at $q=0$ (blue curve).
This implies a stable homogeneous state $\omega(q=0) <0$, but that the large scale instability is faster than regular Cahn--Hilliard dynamics since $\omega(q \rightarrow 0) \neq 0$, which suggests that the passive exchange leads to faster coarsening dynamics.
For $\alpha =1$, we observe that a band of modes is unstable (green curve \figref{fig:app_1}A), implying a low-wave-number (large length scale) cut-off in the instability is present. This is a signature of a stable finite size pattern.

\section{Diffusive efflux and internal unbinding flux} \label{app:diff_int}
We here evaluate the effective fluxes given by \Eqsref{eqn:flux_diff} and \Eqref{eqn:flux_int} in the main text in limiting cases.
For the diffusive efflux~$\Jdiff$ we find
\begin{subequations}
\begin{align}
\Jdiff (R \ll \ell_\mathrm{out}) & =- \frac{2 \pi D_\mathrm{out}}{\gamma+\log(\frac{R}{2\ell_\mathrm{out}})} \Bigl(\phi_\mathrm{out}^\mathrm{eq}(R) - \phi_\mathrm{out}^{(0)}\Bigr) \\
\Jdiff (R \gg \ell_\mathrm{out}) & = \frac{2 \pi R D_\mathrm{out}}{\ell_\mathrm{out}} \Bigl(\phi_\mathrm{out}^\mathrm{eq}(R) - \phi_\mathrm{out}^{(0)}\Bigr) \;.
\end{align}
\end{subequations}
For $R \ll \ell_\mathrm{out}$, we recover the diffusive regime, whereas we recover a flux proportional to the interface (and $D_\mathrm{out}/\ell_\mathrm{out}$) for $R \gg \ell_\mathrm{out}$.
Equating the two limits, we predict a transition between the two regimes at $R^* = -\ell_\mathrm{out}/ W_{-1}(-e^{\gamma/2}) $, where $W(z)$ is the Lambert $W$ function obeying $z=We^W$, and $\gamma$ is Euler's constant. 
For the internal unbinding flux $\Jint$, we find
\begin{subequations}
\begin{align}
\Jint (R \ll \ell_\mathrm{in}) & =\frac{ \pi R^2 D_\mathrm{in}}{\ell_\mathrm{in}^2} \Bigl(\phi^\mathrm{eq}_{\mathrm{in}}(R) - \phi^{(0)}_\mathrm{in}\Bigr) \\
\Jint (R \gg \ell_\mathrm{in}) & = \frac{2 \pi R D_\mathrm{in}}{\ell_\mathrm{in}} \Bigl(\phi^\mathrm{eq}_{\mathrm{in}}(R) - \phi^{(0)}_\mathrm{in}\Bigr) \;,
\end{align}
\end{subequations}
showing that the limit $R \ll \ell_\mathrm{in}$ gives an area-limited flux, and the limit $R \gg \ell_\mathrm{in}$ gives an interface-limited flux. Equating the two limits, we predict a transition at $R^* = 2 \ell_\mathrm{in} $. 

We next estimate the dynamics of the droplet radius in the limiting cases of fast and slow binding. 
The boundary conditions at the droplet surface are fixed due to Laplace pressure, resulting in the Gibbs-Thomson relation $ \phi_n^\mathrm{eq}(R) = \phi_n^{(0)} ( 1+ \ellc/R)$~\cite {Zwicker_2025}.
Here, we assumed identical capillary lengthscales~$\ellc$ in both phases for simplicity.
In the case of fast binding ($R \gg \ell_n$), both fluxes are interface-limited. 
From $ \partial (\pi R^2)/\partial t \propto \Jdiff +\Jint$, we find $ R(t) \propto ( \ellc t [D_\mathrm{out} \phi_\mathrm{out}^{(0)} / \ell_\mathrm{out} + D_\mathrm{in} \phi_\mathrm{in}^{(0)} / \ell_\mathrm{in}])^{1/2}$, leading to the $R \sim t^{1/2}$ scaling, in agreement with previous works~\cite {Wagner1961, Lee_2021}.
In the case of slow binding ($R \ll \ell_n$), $\Jdiff$ is purely diffusive and $\Jint$ is area-limited, leading to different behavior, which we analyze separately assuming that one of the mechanisms dominates transport.
Purely diffusive flux leads to $ R(t) \propto (D_\mathrm{out} \phi_\mathrm{out}^{(0)} \ellc t)^{1/3}$, consistent with Lifshitz--Slyozov theory in two dimensions~\cite {Lifshitz_1961}. 
For dominating area flux, we find $R(t) \propto D_\mathrm{in} \phi_\mathrm{in}^{(0)} \ellc t/ (2 \ell_\mathrm{in})$, leading to linear scaling.
Effective simulations confirm the scalings we predict from this analysis (\figref{fig:app_1}B).
See \appref{app:numerical_details} for numerical details.

In the active case described by \Eqref{eq:effective_droplet_active} in the main text, the effective fluxes become
\begin{subequations}
\begin{align}
\Jdiff^\mathrm{act}  & =  2 \pi  R  \frac{D_\mathrm{out}}{\ell_\mathrm{out}}\frac{ K_1(R/\ell_\mathrm{out})}{K_0(R/\ell_\mathrm{out})} \Bigl(\phi_\mathrm{out}^\mathrm{eq}(R) - \phi_\mathrm{out}^{(0)}- \Delta\phi_\mathrm{out}\Bigr)\;, \\
\Jint^\mathrm{act} & = 2 \pi R \frac{D_\mathrm{in}}{\ell_\mathrm{in}} \frac{I_1(R/\ell_\mathrm{in})}{I_0(R/\ell_\mathrm{in})} \Bigl(\phi^\mathrm{eq}_{\mathrm{in}}(R) - \phi^{(0)}_\mathrm{in} -\Delta\phi_\mathrm{in}\Bigr) \;.
\end{align}
\end{subequations}
The estimated transitions between the different regimes are left unchanged since activity does not modify the $\ell$-dependent terms in the fluxes.
We next estimate again the scaling behaviours in the active case for the enhanced coarsening regime $\Delta \phi_\mathrm{in}>0$, $\Delta \phi_\mathrm{out}<0$.
For small droplets $R \ll l_\mathrm{cap} \phi_n^{(0)}/\Delta \phi_n$, fluxes are unchanged, so we predict the same scalings as in the passive case.
In contrast, for large droplets $R \gg l_\mathrm{cap} \phi_n^{(0)}/\Delta \phi_n$, fluxes are dominated by the active term, so we predict modified scalings. 
For fast exchange ($R \gg \ell_n$), both fluxes are interface-limited, leading to a ballistic scaling $R(t) \sim t$.
For slow exchange ($R \ll \ell_n$), the diffusive flux $\Jdiff$ becomes $R \sim t^{1/2}$, while the internal one becomes exponential $R \sim e^t$.
This regime might be difficult to access, since the modified scalings require large droplets, so it requires $ l_\mathrm{cap} \phi_n^{(0)}/\Delta \phi_n \ll R \ll \ell_n$.
However, we typically consider small activity and strong phase separation, so $\Delta \phi_n \ll 1$, implying that the radius at which the activity starts dominating is large.
Indeed, we observe scalings consistent with the passive fluxes in our numerical simulations (\figref{fig:enhanced_coarsening}A in the main text). 
Simulations with large activity $\Delta \mu =5 \, \kBT$ show that a scaling larger than $R(t) \sim t$ can also be accessed (\figref{fig:app_1}C).

\section{Timescales estimation}\label{app:timescales_estimation}
\subsection{Growth of a single droplet}
To estimate the timescale for the growth of a single droplet, we consider a finite system, in which the final macroscopic droplet is a stationary state of the dynamics.
We use the perturbation decay rate around this stationary state to obtain a characteristic time scale of droplet growth.
Consequently, we expand the equations around the equilibrium concentrations $\phi_n^{\mathrm{eq}}$ and $\psi^\mathrm{eq}$, which obey $s(\phi_\mathrm{n}^{\mathrm{eq}}, \psi^\mathrm{eq})=0$.
We consider the dynamics of a single droplet,
\begin{equation} \label{app:eq:eff_drop}
	\partial_t \phis_{n} \simeq D_{n} \nabla^2 \phis_{n} + \partial_{\phi_n} s(\phi_n^{\mathrm{eq}}) \delta \phi_n + \partial_\psi s(\psi^{\mathrm{eq}}) \delta \psi
	\;,
\end{equation}
where we now cannot neglect the bulk perturbation since the bulk evolves during the dynamics.
Here, we have defined the effective binding kinetics $k_n = -\partial_{\phi_n} s(\phi_n^{\mathrm{eq}})$.
To evaluate the variation of the binding flux with respect to the equilibrium bulk concentration in terms of the binding kinetics, we impose the conservation law for the average fractions,
\begin{equation}
\frac{A(t)}{A_\mathrm{s}} \bar{\phi}_\mathrm{in}(t) + \Bigl( 1- \frac{A(t)}{A_\mathrm{s}} \Bigr) \bar{\phi}_\mathrm{out}(t) + \eta \psi(t) = \PhiTot \;,
\end{equation}
where $A(t)$ represents the area of the droplet, $A_\mathrm{s}$ the area of the surface and $\bar{\phi}_\mathrm{in}(t)$, $\bar{\phi}_\mathrm{out}(t)$ the average fractions in the dense and dilute phases at time $t$.
We can then use $\partial s/ \partial \psi = \partial s/\partial \phi_n \cdot  \partial \phi_n/\partial \psi$ together with approximating the variation of the inside and outside fractions by the variation of their averages,
\begin{align}
\frac{\partial \phi_\mathrm{in}}{\partial \psi} \Big|_\mathrm{eq} &= - \frac{ \eta A_\mathrm{s}}{A_\mathrm{eq}} 
&
\frac{\partial \phi_\mathrm{out}}{\partial \psi} \Big|_\mathrm{eq} &= - \frac{ \eta A_\mathrm{s}}{A_\mathrm{s} - A_\mathrm{eq}} \;,
\end{align}
where $A_\mathrm{eq} = \pi R_\mathrm{eq}^2$ represents the equilibrium volume of the droplet.
Inserting this in \Eqref{app:eq:eff_drop}, we obtain the same functional form as \Eqref{appA:eq:eff_dyn} with the time-dependent binding equilbrium
\begin{align}
\phi_\mathrm{in}^{(0)}(t) & = \phi_\mathrm{in}^\mathrm{eq} + \frac{A_\mathrm{s}}{A_\mathrm{eq}} \eta ( \psi(t) - \psi_\mathrm{eq} )
&
\phi_\mathrm{out}^{(0)}(t) & = \phi_\mathrm{out}^\mathrm{eq} + \frac{A_\mathrm{s}}{A_\mathrm{s} - A_\mathrm{eq}} \eta ( \psi(t) - \psi_\mathrm{eq} ) \;.
\end{align}
We then assume that the motion of the interface is slower than the equilibration of the concentration profile, and solve the Helmolz equations in the two phases at stationarity.
For the dense phase, we impose the boundary conditions $\phi_{\mathrm{in}}(R) = \phi_\mathrm{in}^\mathrm{eq}(R)$ and $\phi_{\mathrm{in}}'(r=0)=0$.
For the dilute phase, we consider boundary conditions $\phi_{\mathrm{out}}(R) = \phi_\mathrm{out}^\mathrm{eq}(R)$ and $\phi_{\mathrm{out}}(L )=\phi_\mathrm{dil}(t)$.
Taken together, we find
\begin{subequations} \label{eq:single_drop_dynamics_appendix}
\begin{align}
	\partial_t A(t) & = \frac{1}{\phi_\mathrm{in}^{(0)} - \phi_\mathrm{dil}(t)} \Bigl( - J_\mathrm{diff} - J_\mathrm{int} \Bigr)
\\
	\partial_t \phi_\mathrm{dil}(t) & = \frac{1}{A_\mathrm{s} - A(t)} \Bigl( J_\mathrm{diff} - J_\mathrm{ext} \Bigr) 
 \\
	\partial_t \psi(t) & = \frac{\nub }{V_\mathrm{b} \nus} \Bigl( J_\mathrm{int} + J_\mathrm{ext} \Bigr) 
	\;,
\end{align}
\end{subequations}
where $\phi_\mathrm{dil}(t) $ is the volume fraction on the surface far away from the droplet, $\psi(t)$ denotes the volume fraction in the bulk, and $V_\mathrm{b}$ the volume of the bulk.
The exchange between the compartments is given by the diffusive efflux $J_\mathrm{diff}$, which quantifies the amount of material leaving the droplet on the surface via diffusion through its interface, and the internal and external unbinding fluxes $J_\mathrm{int}$ and $J_\mathrm{ext}$, respectively quantifing the amount of the material going from the dense and dilute phases to the bulk.
 These can be determined using the profiles calculated above,
\begin{subequations} \label{eq:3fluxes}
\begin{align}
J_\mathrm{diff} & = - 2 \pi R D_\mathrm{out} \Bigl( \partial_r \phi_\mathrm{out}(r) |_R  \Bigr)
\\
J_\mathrm{int} & = -2 \pi  \int_{0}^{R} r (\phi_\mathrm{in}(r,t) - \phi_\mathrm{in}^{(0)}(t)) \mathrm{d}r, 
\\
J_\mathrm{ext} & = -2 \pi  \int_{R}^{L} r (\phi_\mathrm{out}(r,t) - \phi_\mathrm{out}^{(0)}(t)) \mathrm{d}r \;,
\end{align}
\end{subequations}
upon expanding the binding flux as above. The larger eigenvalue of the Jacobian of \Eqsref{eq:single_drop_dynamics_appendix} evaluated at the large droplet stationary state $(A_\mathrm{eq}, \phi_\mathrm{out}^\mathrm{eq},\psi^\mathrm{eq})$
determine the timescales shown in the blue curve in \figref{fig:passive_coarsening}E in the main text.

\subsection{Coarsening timescale}
To evaluate the timescale for coarsening, we assume the overall binding equilibrium between the surface and bulk has been reached, and any dynamics is exclusively due to material exchange between droplets (potentially mediated by the bulk).
In particular, we study the dynamics of a droplet of radius $R_1$ surrounded by stable droplets of radius $R_2$ at distance $L$.
We can then determine the concentration profiles considering again that the motion of the interface is slower than the equilibration of the concentration profiles, and solving \Eqref{appA:eq:eff_dyn} quasi-statically.
Such droplets are in equilibrium with the bulk, implying $\phi_\mathrm{out}^{(0)} = \phi_\mathrm{out}^{\mathrm{eq}}(R_2)$, such that they change only in response to the dynamics of the first droplet, and are otherwise stationary.
The associated boundary conditions in the dilute phase are $\phi_\mathrm{out}(R_1) = \phi_\mathrm{out}^{\mathrm{eq}}(R_1)$  and $\phi_\mathrm{out}(L) = \phi_\mathrm{out}^{\mathrm{eq}}(R_2)$. 
From the resulting concentration profile, we can determine the diffusive flux $\Jdiff^{\mathrm{coars}}=-2 \pi R_1  D_\mathrm{out}  \phi_{\mathrm{out}}'(R_1)$ as above, which now depends on the distance $L$ between droplets and the second droplet's radius $R_2$. 
Similarly, to determine the dynamics due to the internal flux, we impose that the overall binding equilibrium was already reached internally, so $\phi_\mathrm{in}^{(0)} = \phi_\mathrm{in}^\mathrm{eq}(R_2)$. 
Inserting this in \Eqref{eqn:flux_int} of the main text, we determine the internal flux, which now depends on $R_2$.
The evolution of the first droplet with radius $R_1$ then reads 
\begin{equation} \label{eq:app:eff_R_dyn}
\partial_t R_1  = \frac{1}{2 \pi R_1 (\phi_\mathrm{in}^{(0)} - \phi_\mathrm{out}^{(0)}) } \Bigl( -\Jdiff^{\mathrm{coars}} (R_1,R_2) - \Jint(R_1,R_2) \Bigr) \;.
\end{equation}
We consider the solution in which all droplet have the same radius $R_1=R_2=R_*$ and consider the evolution of small perturbations using linear stability analysis.
This perturbation grows exponentially with a timescale $T_\mathrm{coars}=1/(\partial_{R_1}  f(R_1,R_2)|_{R_1=R_2=R_*})$, determining the orange curve shown in \figref{fig:passive_coarsening}E in the main text.

\subsection{Active coarsening timescale} \label{app:act_coars_time}
We next repeat the calculation above including the active corrections $\Delta \phi_\mathrm{n}$ introduced in \Eqref{eq:effective_droplet_active} in the main text.
In this case, the stationary configuration considered above with all droplets of the same size becomes dependent on the exchange kinetics $k$, and thus on the reaction diffusion lengthscale $\ell$. 
This implies we cannot calculate the timescale depending on $\ell$ as done above, since the stationary state now also depends on it, so we consider a particlar case.
Focusing on buddying yeast, we are interested in estimating the activity needed for a speed-up that makes the coarsening timescale consistent with the experimentally measured one.
We calculate the active corrections $\Delta \phi_\mathrm{n} = -s(\phizn)/s'(\phizn)$ using \Eqref{eq:app:deltaphis}, assuming small activities.
We here consider the case of active unbinding from dilute regions ($\alpha=-1$), which leads to accelerated coarsening. 
We calculate the derivatives as $s'(\phi_\mathrm{in}^{(0)}) = - k f_\mathrm{s}''(\phi_\mathrm{in}^{(0)})$, $s'(\phi_\mathrm{out}^{(0)}) = - 2 k f_\mathrm{s}''(\phi_\mathrm{out}^{(0)})$.
Moreover, to calculate the basal fluxes we use that their sign is determined by the active part (\appref{app:eff_drop}), so we consider the simple case that this term dominates, %
\begin{align}
	s_\mathrm{in}(\phi_\mathrm{in}^{(0)}) & \sim  k_\mathrm{p} \Bigl( \frac{ \Delta \mu } {2 \kBT} \Bigr)
	&
	s_\mathrm{out}(\phi_\mathrm{out}^{(0)}) & \sim  k_\mathrm{p} \Bigl(  -\frac{\Delta \mu } {2 \kBT} \Bigr)
	\;.
\end{align}
We choose $\chi=3.5$, $\phi_\mathrm{in}^{(0)}=0.95$, $\phi_\mathrm{out}^{(0)} = 0.035$, and $\Delta \mu = 2 \kBT$, resulting in $\Delta \phi_\mathrm{in} \simeq 0.07$ and $\Delta \phi_\mathrm{out} \simeq -0.02$.
Inserting this in the equation for the dynamics of the radius given by \Eqref{eq:app:eff_R_dyn} assuming $R_1 = R_2$ and correcting for $\Delta \phi_\mathrm{n}$, we find a stationary radius $R_* \sim 100\,l_\mathrm{c}$, consistent with the size of the polarity spot.
The system relaxes back to such stationary state exponentially, with a timescale $T_\mathrm{coars}^\mathrm{act} \sim \SI{100}{\second}$.

\begin{figure}[t]
  \centering
    {\includegraphics[width= .7\columnwidth]{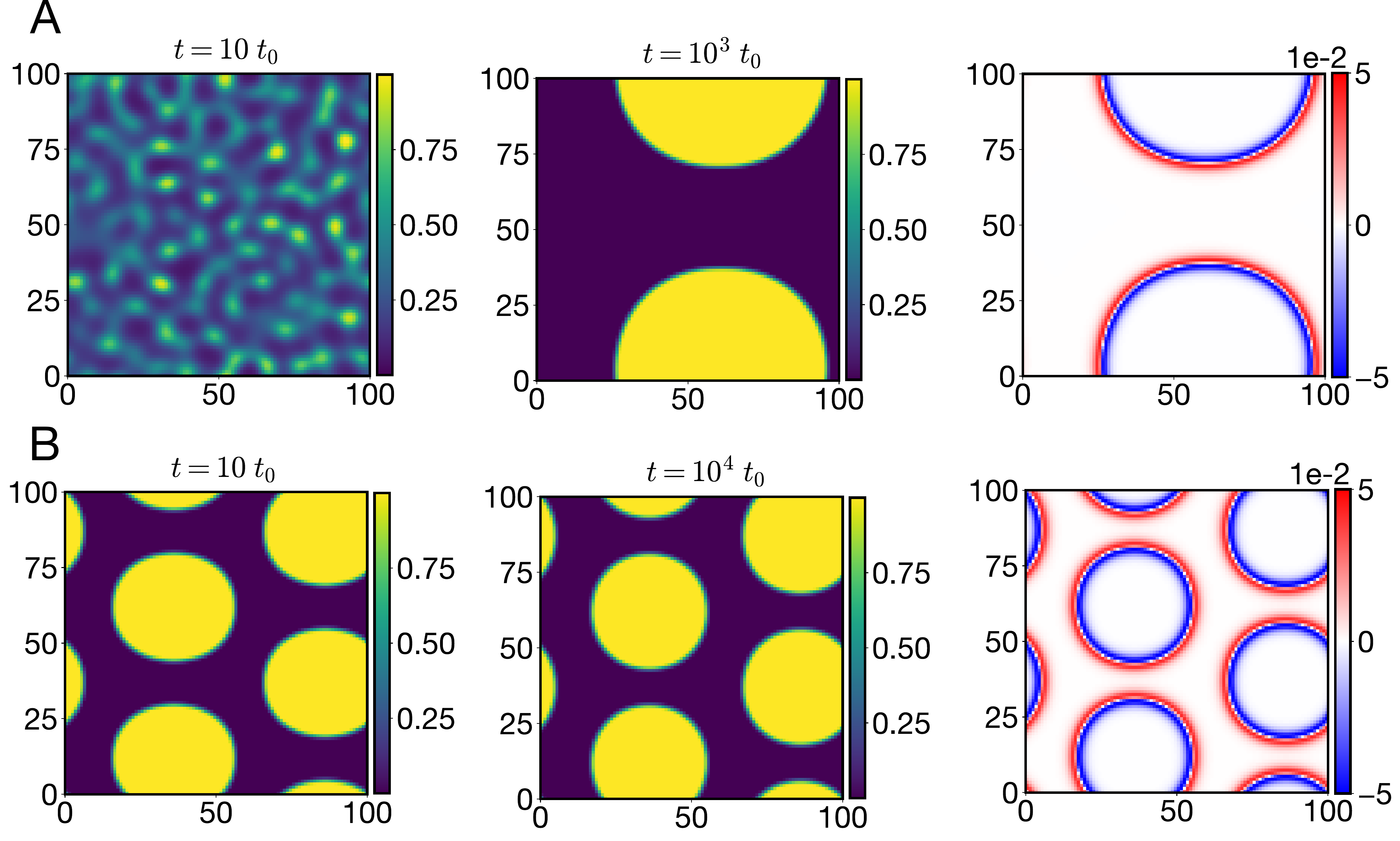}}
  \caption{ \textbf{Active unbinding from dense regions exhibits multistability}
   (A) Initial condition given by perturbing homogeneous state that is a zero of the homogeneous binding flux.
   (B) Inital condition with multiple droplets.
   Left panels: concentration profiles at early times. Central panel: concentration profiles at stationarity.
   Right panel: corresponding exchange fluxes at stationarity.
   Remaining parameters are the same as \figref{fig:active_binding}D in the main text.
}
\label{fig:app_multi}
\end{figure}

\section{Multi-Stability}
We performed numerical simulations that demonstrate multistability in the regime associated with the regular coarsening in the case of fast exchange with active unbinding from dense regions (yellow curve \figref{fig:active_binding}A in the main text).
Initialising the system in the homogeneous binding equilibrium (zeros of \Eqref{eq:active_binding_flux} in the main text) and perturbing it leads to coarsening (\figref{fig:app_multi}A).
In contrast, initializing the system with multiple droplets leads to a stationary state with multilple droplets of the same size (\figref{fig:app_multi}B), similarly to what happens for intermediate exchange rates (pink curve \figref{fig:active_binding}A in the main text) and area-limited droplets.

\section{Numerical details} \label{app:numerical_details}
All simulations in the main text start are done using a finite difference method implemented in the Python package \textit{py-pde}~\cite {Zwicker_2020}.
The initial state is determined by a random perturbation of a homogeneous surface state satisfying binding equilibrium (i.e., zeros of \Eqref{eq:active_binding_flux} in the main text). 
We use simulations using a thin-interface approximation~\cite {Kulkarni2023} to run simulations for a long time and get the scalings associated with different transport mechanisms shown in \figref{fig:app_1}B.

\end{document}